\documentclass{webofc}
\usepackage[varg]{txfonts}   % Web of Conferences font
\usepackage{hyperref}
\usepackage{url}
\hypersetup{colorlinks=true,citecolor=blue,urlcolor=blue,linkcolor=blue}
\begin{document}
\title{Equilibrium expectations for non-Gaussian fluctuations near a QCD critical point}

\author{\firstname{Jamie M.} \lastname{Karthein}\inst{1}\fnsep\thanks{\email{jmkar@mit.edu}} \and
        \firstname{Maneesha} \lastname{Pradeep}\inst{2} \and
        \firstname{Krishna} \lastname{Rajagopal}\inst{1}
        \and
        \firstname{Mikhail} \lastname{Stephanov}\inst{3}
        \and
        \firstname{Yi}
        \lastname{Yin}\inst{4}
}

\institute{Center for Theoretical Physics, Massachusetts Institute of Technology, Cambridge, MA 02139,
USA 
\and
Department of Physics, University of Maryland, College Park, Maryland 20742, USA
\and
Department of Physics, University of Illinois, Chicago, Illinois 60607, USA
\and
School of Science and Engineering, The Chinese University of Hong Kong, Shenzhen, Guangdong, 518172, China
          }

\fnsep{\!\!\!\!\!MIT-CTP/5662}

\abstract{
%Abstract modified from time of submission
With the highly anticipated results from the Beam Energy Scan II program at RHIC being recently revealed, an understanding of particle-number fluctuations and their significance as a potential signature of a possible QCD critical point is crucial. 
Early works that embarked on this endeavor sought to estimate the fluctuations due to the presence of a critical point assuming they stay in equilibrium. 
From these results came the proposal to focus efforts on higher, non-Gaussian, moments of the event-by-event distributions, in particular of the number of protons. 
These non-Gaussian moments are especially sensitive to critical fluctuations, as their magnitudes are proportional to high powers of the critical correlation length. 
As the equation of state provides key input for hydrodynamical simulations of heavy-ion collisions, we estimate equilibrium fluctuations from the BEST equation of state (EoS) that includes critical features from the 3D Ising Model. 
In particular, the proton factorial cumulants and their dependence on non-universal mapping parameters is investigated within the BEST EoS.
Furthermore, the correlation length, as a central quantity for the assessment of fluctuations in the vicinity of a critical point, is also calculated in a consistent manner with the scaling equation of state. 
An understanding of the equilibrium estimates of proton factorial cumulants will be useful for further comparison to estimates of out-of-equilibrium fluctuations in order to determine the magnitude of the observable fluctuations to be expected in heavy-ion collision experiments, in which the time spent near a critical point is short.
}
\maketitle
\section{Introduction}
\label{intro}
The phase structure of QCD matter at finite temperature and density remains an open question in nuclear science.
While first-principles lattice QCD calculations are well-suited for vanishing density, finite density results are limited to $\mu_B/T \sim 3.5$ \cite{Borsanyi:2021sxv}.
Experimental measurements are, therefore, the only modern tool that can access further reaches of the phase diagram.
The Beam Energy Scan II (BESII) program at RHIC, which had the goal of mapping out the phase diagram, recently completed its runs and revealed measurements of the proton fluctuations for a range of $\mu_B$~\cite{Pandav:CPOD2024}.
The moments of the proton distribution are sensitive to the presence of a critical point in the phase diagram with higher orders diverging more strongly due to higher powers of the correlation length, a quantity which diverges at the critical point.
(For a review see Ref. \cite{Bzdak:2019pkr}.)
These results show no evidence of a critical point in the region of the phase diagram with $\mu_B \lesssim 420~{\rm MeV}$ where experimental measurements have focused to date.
However, theoretical models of the noncritical baseline expected for these fluctuations do not appear to describe the data at all energies. 
Therefore, it is crucial to understand the critical point contribution to proton fluctuations in order to help interpret these results.

\section{Scaling Equation of State Mapped to QCD}
\label{sec:method}

Studies of the effect of a critical point on the equation of state for QCD have been underway with the advent of the BEST equation of state in the original work of Parotto et al \cite{Parotto:2018pwx}.
This equation of state incorporates critical features based on the universality principle of scaling near a critical point.
The universality class for QCD has been shown to be that of the 3D Ising model, as discussed in the review Ref. \cite{Bzdak:2019pkr}.
The 3D Ising model scaling equation of state is given by the parametric form:
\begin{equation}
\label{GFreeEner}
\begin{split}
     P_{\text{Ising}} &= - G(R,\theta)
     = h_0 M_0 R^{2 - \alpha}(\theta \tilde{h}(\theta) - g(\theta)),
\end{split}
\end{equation}
where $h_0, M_0$ are normalization constants for the magnetic field and magnetization, $\alpha=0.11$ is a 3D Ising critical exponent, and $\tilde{h}(\theta)$ and $g(\theta)$ are polynomials in $\theta$ \cite{Parotto:2018pwx,Karthein:2021nxe}.
The critical fluctuations are calculated as derivatives of the equation of state.
For example the n-point correlations of baryon density are given by the $k^{th}$ derivative of pressure with respect to baryon chemical potential, $\mu_B$ at fixed temperature:
\begin{equation}
\langle(\delta n)^{k}\rangle=\frac{T^{k-1}}{V^{k-1}}\frac{\partial^{k}P}{\partial\mu^{k}}
\end{equation}

Higher order fluctuations diverge more strongly at the critical point due to their scaling with higher powers of the correlation length.
The correlation length itself may also be written the parametric form in the Ising model in terms of the parametric variables $(R,\theta)$ as \cite{ZinnJustin}:
\begin{equation}
\label{ZinnJustin}
    \xi^2(M,t) = R^{-2 \nu} x_\xi(\theta).
\end{equation}
We present the $\theta$-dependence of the correlation length as calculated in the $\epsilon$-expansion to $\mathcal{O}(\epsilon^2)$:
\begin{equation}
\label{xiOeps2}
    x_\xi = x_\xi (0)\Bigg(1-\frac{5}{18}\epsilon \theta^2 + \Big[\frac{1}{972}( 24 I-25)\theta^2 + \frac{1}{324} ( 4 I + 41) \theta^4\Big] \epsilon^2)\Bigg),
\end{equation}
where $I \equiv \int_0^1 \frac{\ln [x (1-x)]}{1 - x(1-x)}dx \approx -2.3439$.
This updated correlation length at $\mathcal{O}(\epsilon^2)$ is, furthermore, consistent with the scaling equation of state valid to the same order in the $\epsilon$-expansion.

In order to map the critical features of the 3D Ising model onto the QCD phase diagram, we employ a quadratic mapping as first introduced in Ref. \cite{Kahangirwe:2024cny}.
This is an improvement over the original work of the BEST collaboration where a linear mapping allowed for the Ising transition line to be mapped onto the tangent to the chiral phase transition line from lattice QCD.
This quadratic mapping is combined with an alternative expansion scheme, known as the  $T'$-expansion, introduced in \cite{Borsanyi:2021sxv}, which is a essentially a reshuffling of the Taylor expansion.
The QCD variables $\mu_B^{2}-\mu_{B,c}^2$ and $T'(\mu,T)-T_0$, are related to the Ising variables $(r,h)$ near the critical point as:
\begin{equation}
\label{eq:quadmap}
    \frac{T'(\mu_B,T)-T_0}{T^{'}_{T}T_c}
    = - h w \frac{\sin(\alpha_{1}-\alpha_2)}{\cos\alpha_1}\,;\quad
    \frac{\mu_B^2-\mu^2_{B,c}}{2\mu_{B,c} T_c}
    =-w(r\rho\cos\alpha_1
    +h\cos\alpha_2)\,.
\end{equation}
where the mapping parameters $T_c$, $\mu_{B,c}$, $\alpha_{1,2}$, $w$ and $\rho$ were introduced in Ref.\cite{Parotto:2018pwx}.
The location of the critical point in the phase diagram is given by $(\mu_{B,c}, T_c)$ while the size and shape of the critical region are controlled by the remaining parameters.
However, by fixing the critical point to lie along the chiral phase transition line as in Refs. \cite{Parotto:2018pwx,Karthein:2021nxe,Kahangirwe:2024cny}, $T_c$ and $\alpha_1$ are fixed by the choice of $\mu_{B,c}$, leaving 4 free parameters.

In order to calculate the critical contribution to the proton factorial cumulants, $\hat{\Delta}\omega_p^k$, from the fluctuations of the thermodynamic densities we utilize the maximum entropy approach from Ref. \cite{Pradeep:2022eil}.

\section{A Selection of Results, and a Look Ahead}

\begin{figure}
\centering
\begin{minipage}{0.3\textwidth}
  \includegraphics[width=\textwidth]{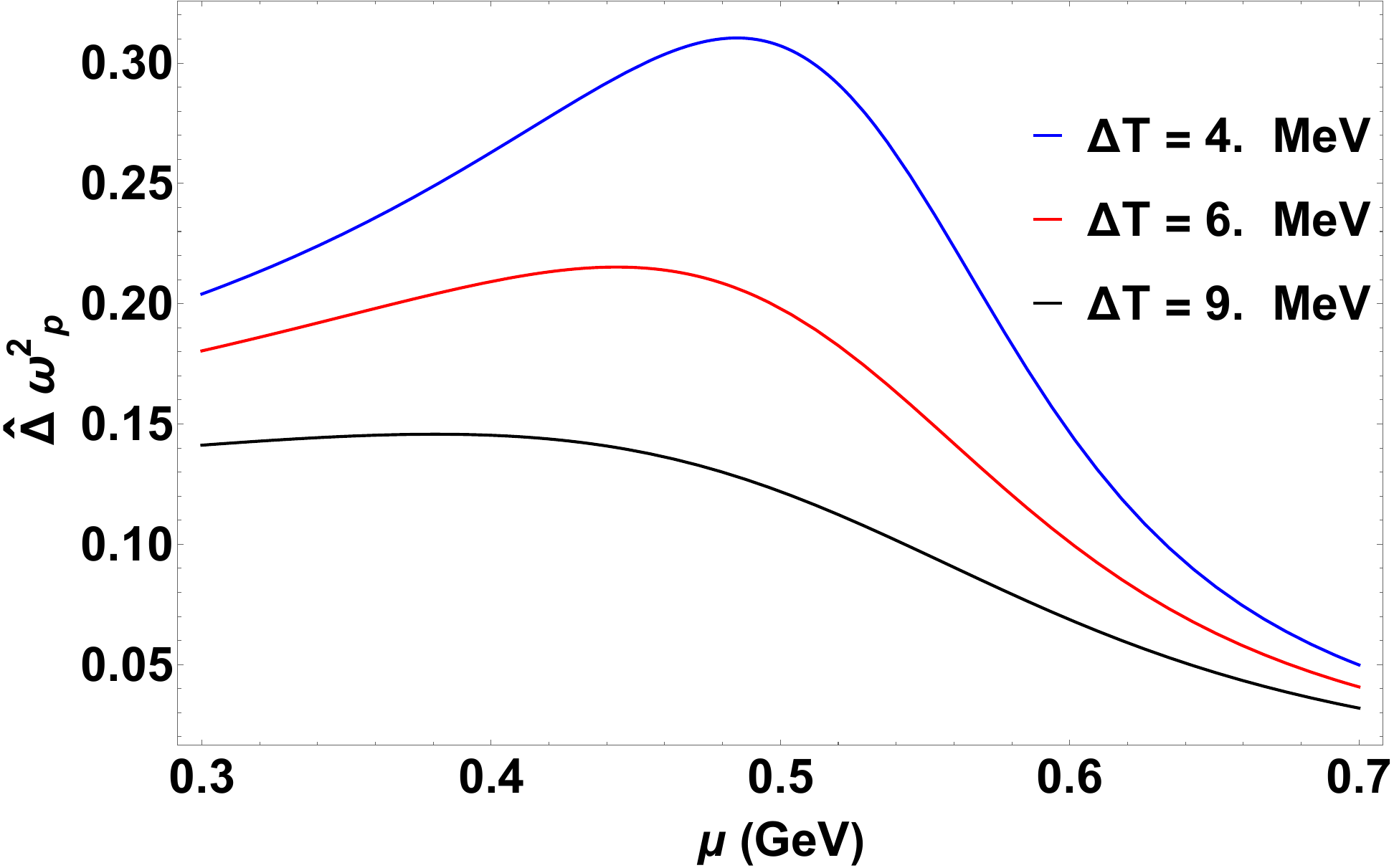}
  \end{minipage}
  \begin{minipage}{0.3\textwidth}
  \includegraphics[width=\textwidth]{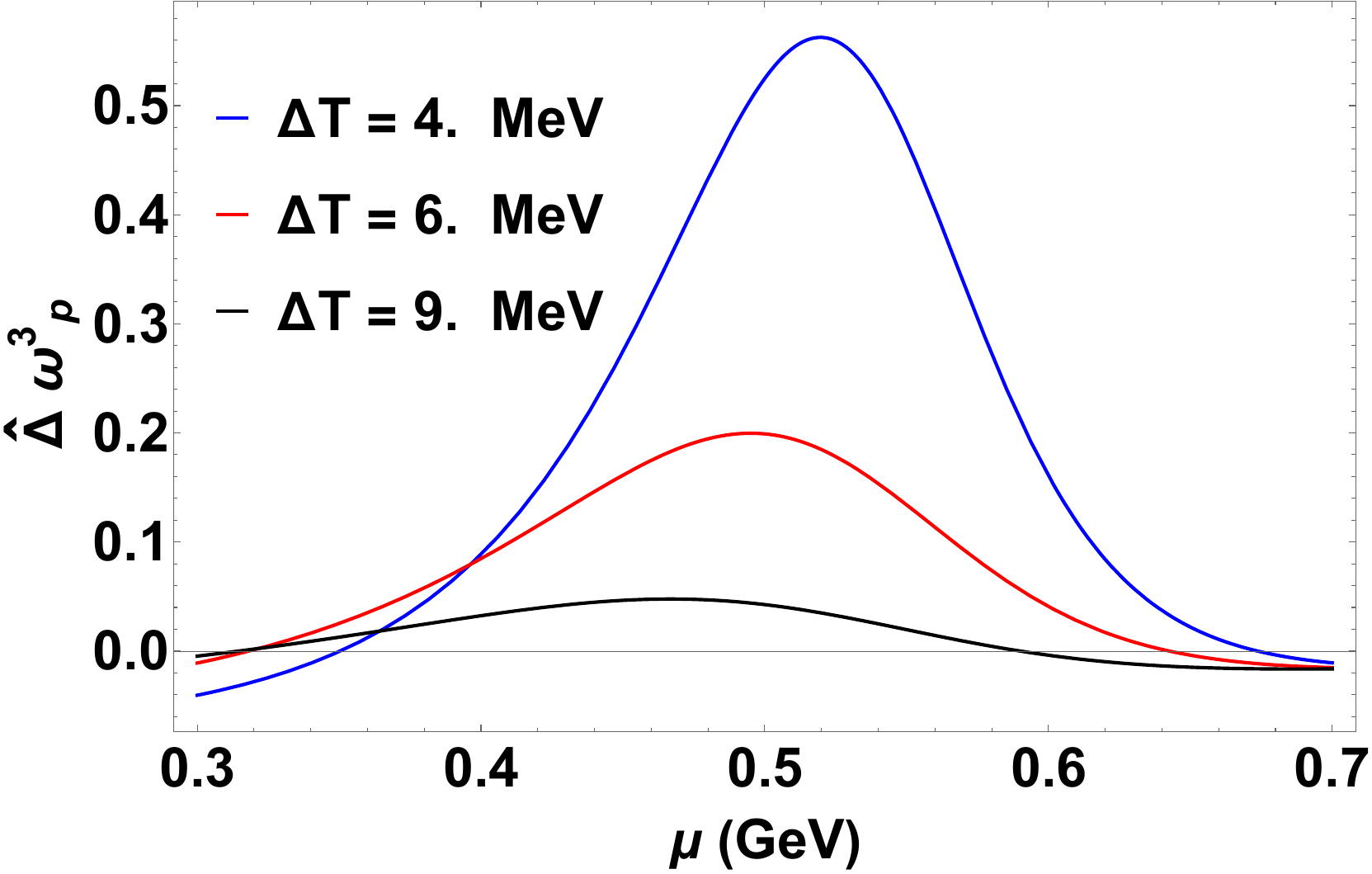}
  \end{minipage}
   \begin{minipage}{0.3\textwidth}
  \includegraphics[width=\textwidth]{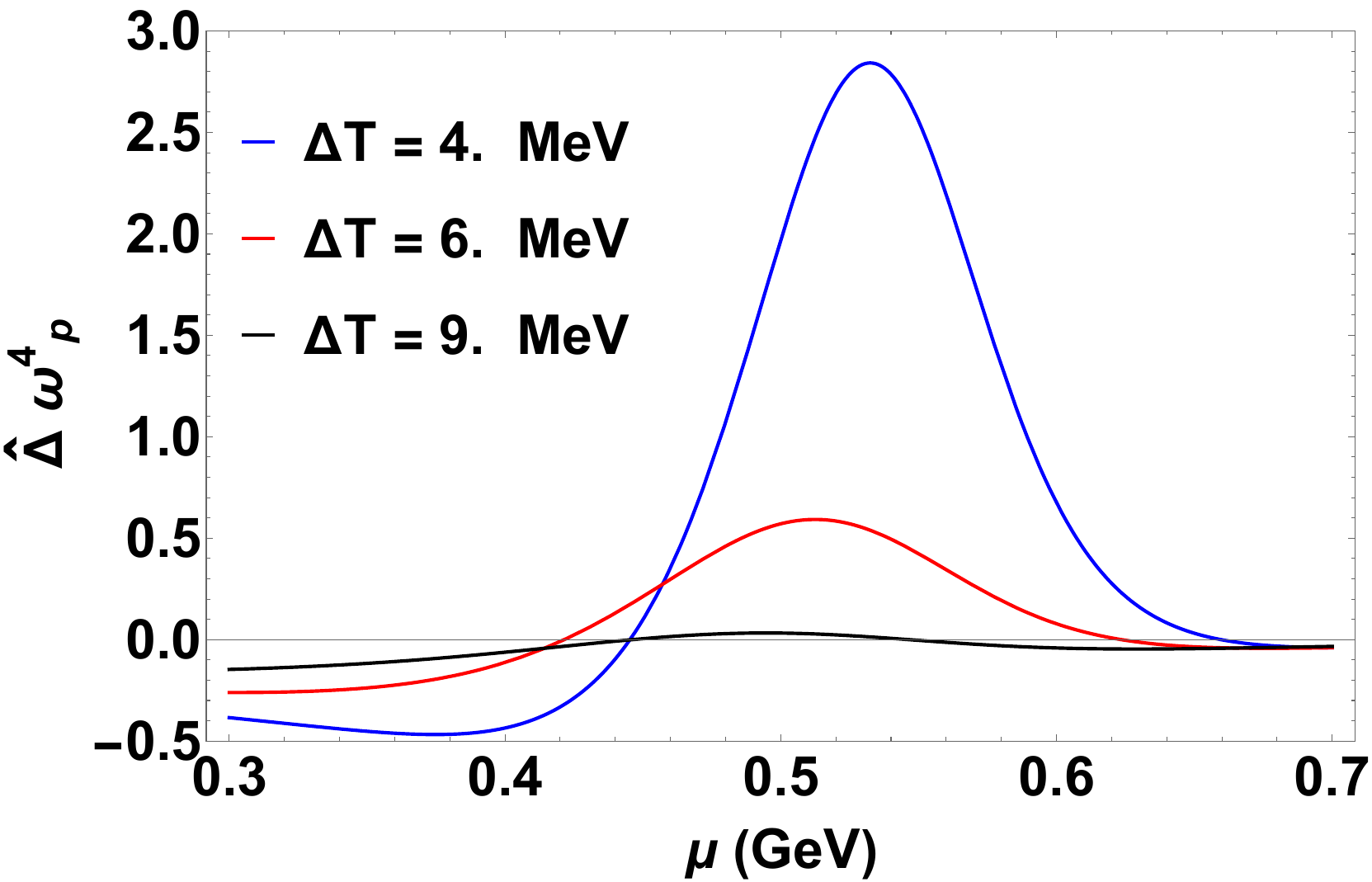}
  \end{minipage}
  \caption{$\hat{\Delta}\omega^k_p$ plotted for $k=2,3,4$ with the choice of $\mu_{B,c}=600, \text{MeV}$,  $\alpha_2=0^{\circ}$, $w=5$ and $\rho=1$ along different freeze-out trajectories characterized by $\Delta T$.}
\label{fig:Freezeout_omegas_w5_rho1_alpha2_0}
\end{figure}

We show the equilibrium results for the critical contribution to the second, third and fourth order fluctuations, $\langle(\delta n)^{k}\rangle$, in the QCD phase diagram for a given choice of the mapping parameters: $\mu_{B,c}=600 ~ \rm{MeV}, ~ w=5, ~ \rho=1, ~ \alpha_2=0$.
Due to the constraint that the critical point should lie along the chiral phase transition line, the choice of $\mu_{B,c}=600$ MeV fixes the remaining two parameters, $T_c=90$ MeV, $\alpha_1=16.6^{\rm{o}}$.
Given this choice, we see that within the critical region the contours for these fluctuations obtain the typical features as expected from Ref. \cite{Bzdak:2019pkr}, including the strengthening of fluctuations toward the critical point, a sign change for the third cumulant from  negative (red) to positive (blue) across the transition line, and a double-lobed structure for the fourth cumulant.
Furthermore, due to the quadratic mapping between Ising and QCD variables as presented in Ref. \cite{Kahangirwe:2024cny} and shown here in Eq. \eqref{eq:quadmap}, we see that these features follow the curve of the chiral phase transition line rather than a line tangent to the critical point as in Refs. \cite{Parotto:2018pwx,Karthein:2021nxe}.
This in turn gives rise to the behavior of the factorial cumulants as shown in Fig. \ref{fig:Freezeout_omegas_w5_rho1_alpha2_0} along freeze-out lines in the phase diagram given by a shift $\Delta T$ from the transition line, which follow curves 4, 6, and 9 MeV below the critical point itself.
Along these freeze-out lines, we see a peak in all three cumulants near the critical point with the magnitude of the peak increasing with increasing order of the fluctuations.
Furthermore, a sign change can be seen in third and fourth cumulants.
As expected, in thermal equilibrium the magnitude of the fluctuations depends strongly on the distance $\Delta T$ between the freeze-out and the critical point, decreasing with increasing $\Delta T$.
We note that the choice of parameters we present here provides an example of what we can expect for equilibrium fluctuations.
The features of these curves as manifest from the size and shape of the critical region are highly dependent on the choice of parameters.
In fact, because the proton factorial cumulants are so sensitive to the parameters, these parameters can be constrained by the experimental measurements.
In a forthcoming work, we will more fully explore the effect of the mapping parameters on the proton factorial cumulants.
Furthermore, so far we have focused on equilibrium results by utilizing the equation of state, however, we will also study the non-equilibrium effects in order to provide a more direct comparison with experiment.

\begin{figure}
\centering
\begin{minipage}{0.3\textwidth}
  \includegraphics[width=\textwidth]{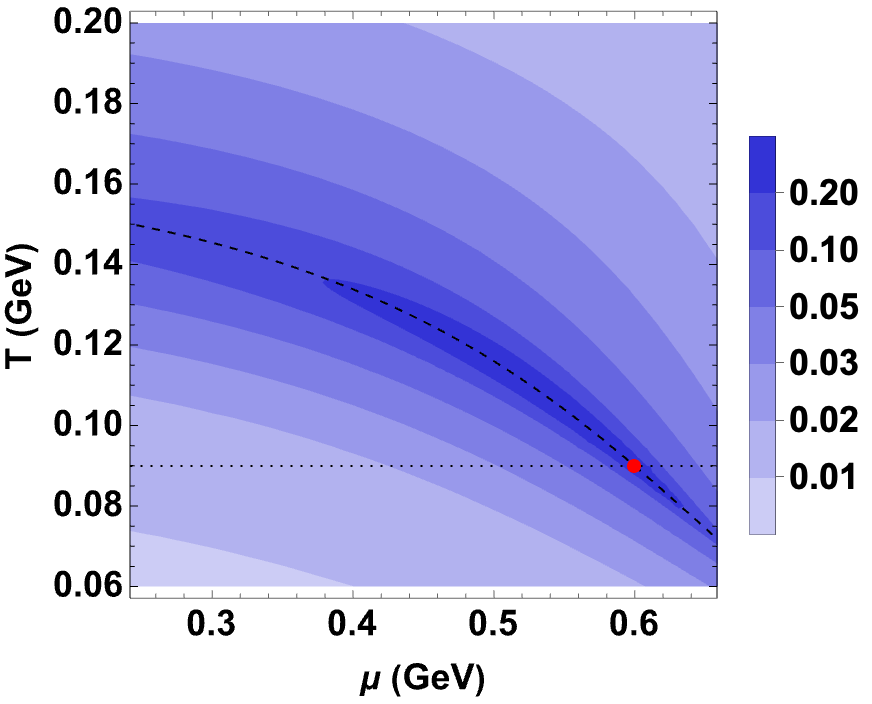}
  \end{minipage}
  \begin{minipage}{0.3\textwidth}
  \includegraphics[width=\textwidth]{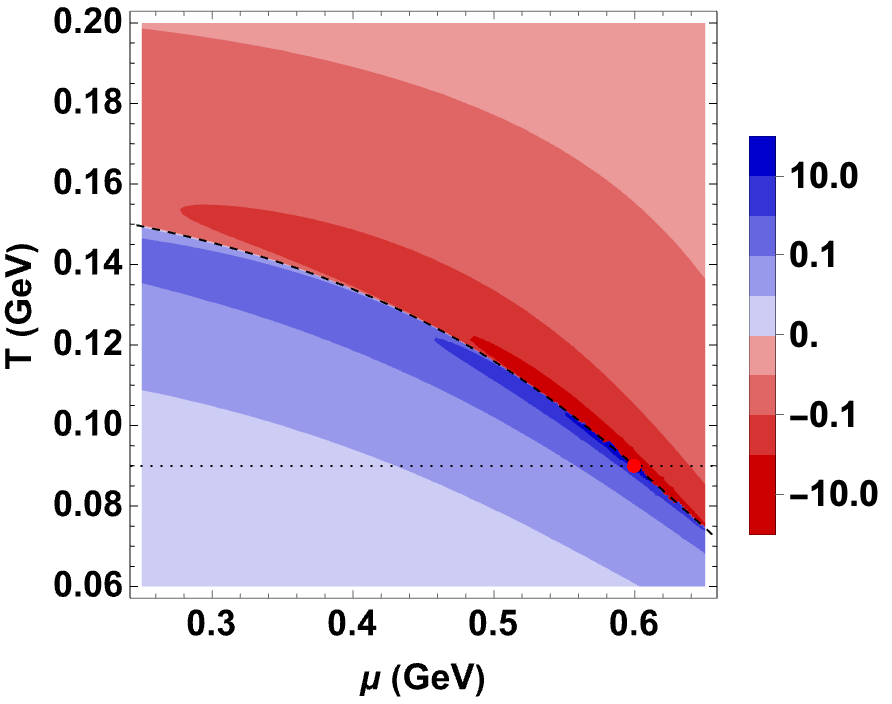}
  \end{minipage}
   \begin{minipage}{0.3\textwidth}
  \includegraphics[width=\textwidth]{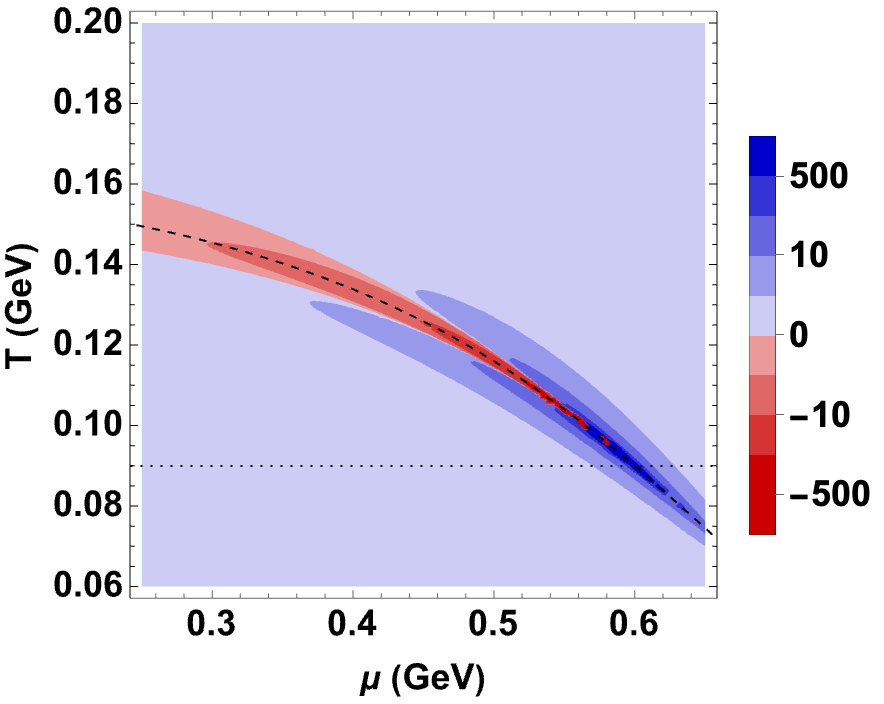}
  \end{minipage}
  \caption{Contours of the critical contribution to $V^{k-1} T_c^{-3}\langle(\delta n)^k\rangle$ for $k=2,3,4$ with the choice of $\mu_{B,c}=600, \text{MeV}$, $\alpha_2=0^{\circ}$, $w=5$ and $\rho=1$}
\label{fig:PhaseDiag_omegas_w5_rho1_alpha2_0}
\end{figure}

\section{Acknowledgements}
JMK is supported by an MPS Ascend Fellowship from the National Science Foundation under Award No. 2138063. This work is supported by the U.S.~Department of Energy, Office of Science, Office of Nuclear Physics grants DE-SC0011090, DE-FG02-01ER41195 and DE‐FG02‐93ER40762.

% BibTeX or Biber users please use (the style is already called in the class, ensure that the "woc.bst" style is in your local directory)
% \bibliography{your_bib_file} % Replace "your_bib_file" with the actual name of your .bib file
\bibliography{all.bib}

\end{document}